# Low-defectiveness exfoliation of MoS$_2$ nanoparticles and their embedment in hybrid light-emitting polymer nanofibers

Alberto Portone,[a,b] Luigi Romano,[a,b] Vito Fasano,[a] Riccardo Di Corato,[a,c] Andrea Camposeo,[b] Filippo Fabbri,[d] Francesco Cardarelli,[e] Dario Pisignano,[b,f] Luana Persano[b]

[a.] *Dipartimento di Matematica e Fisica 'Ennio De Giorgi', Università del Salento, via Arnesano, I-73100 Lecce, Italy*
[b.] *NEST, Istituto Nanoscienze-CNR, Piazza San Silvestro 12, I-56127 Pisa, Italy. E-mail: luana.persano@nano.cnr.it*
[c.] *Center for Biomolecular Nanotechnologies (CBN), Istituto Italiano di Tecnologia, Via Barsanti, 73010 Arnesano (LE), Italy*
[d.] *Center for Nanotechnology Innovation @NEST, Istituto Italiano di Tecnologia, Piazza San Silvestro 12, I-56127 Pisa, Italy*
[e.] *NEST, Scuola Normale Superiore, Piazza San Silvestro 12, I-56127 Pisa, Italy*
[f.] *Dipartimento di Fisica, Università di Pisa, Largo B. Pontecorvo 3, I-56127 Pisa, Italy. E-mail: dario.pisignano@unipi.it*

Molybdenum disulfide (MoS$_2$) has been attracting extraordinary attention for its intriguing optical, electronic and mechanical properties. Here we demonstrate hybrid, organic-inorganic light-emitting nanofibers based on MoS$_2$ nanoparticle dopants obtained through a simple and inexpensive sonication process in *N*-methyl-2-pyrrolidone and successfully encapsulated in polymer filaments. Defectiveness is found to be kept low, and stoichiometry preserved, by the implemented, gentle exfoliation method that allows the MoS$_2$ nanoparticles to be produced. So-achieved hybrid fibers are smooth, uniform, flawless, and exhibit bright and continuous light emission. Moreover, they show significant capability of waveguiding self-emitted light along their longitudinal axis. These findings suggest the use of emissive MoS$_2$ fibers enabled by gentle exfoliation methods as novel and highly promising optical material for building sensing surfaces and as components of photonic circuits.

**1. Introduction**

Layered molybdenum disulfide (MoS$_2$)[1-4] is a quasi-two-dimensional material with covalent in-plane bonds and weak out-of-plane van der Waals interactions. Such transition-metal dichalcogenide shows highly interesting semiconducting properties,[5] including sizable direct band gap (1.2-2.2 eV),[2,6,7] and can be easily intercalated, thus being potentially useful for a variety of electronic applications.[3,8,9] In nanoparticles (NPs), isolated or interspersed within flakes, MoS$_2$ shows quantum confinement effects, with typical blue-shift of photoluminescence (PL) upon crystal size reduction,[7,10] as well as excitation-dependent PL emission.[10,11] These properties, combined with the potentially improved brightness compared to mono- and few-layered forms,[7] make MoS$_2$ NPs highly attractive for the realization of tunable light sources and optical sensors.

Embedding such materials in hybrid polymer architectures, such as microspheres or fibers, would offer additional advantages, enabling the easier manipulation of the functional NPs and their integration in flexible optoelectronic components, and in terms of improved stability.[12-15] For instance, fibers with diameter up to few micrometers, which can be straightforwardly electrospun by intense electric fields (0.01-0.1 MV/m) applied to NP-polymer solutions with sufficient degrees of entanglement,[16-18] would make MoS$_2$ dopants utilizable in photonic circuits and boards, including arrays of optical sensors and synaptic components.[19,20] The process is highly versatile once proper dispersion conditions are found for NPs, leading to either individual filaments or large-area mats consisting of very long and flexible fibers. In previous works, both hybrid and single-component light-emitting materials were processed, showing quantum yield higher than in corresponding conjugated polymer films, chain alignment and optical anisotropy,[21-23] and formation of electrostatically-interacting molecular systems with intercoupled optomechanical properties.[24] The encapsulation of MoS$_2$ flakes in carbon fibers has been previously explored as strategy for the realization of anodes for high-performance batteries,[25,26] however the study of the emission properties of these compounds in polymer nanofibers has not been carried out before, as well as their use to realize prototypical photonic components.

Focusing on optical properties, here we introduce light-emitting fibers based on MoS$_2$ NPs interspersed in flakes, electrospun jointly with a poly(methyl methacrylate) (PMMA) matrix. NPs are obtained through gentle exfoliation, which is found to reduce the MoS$_2$ particle size in all the three spatial dimensions, while concomitantly preserving low defect concentration and stoichiometry as highlighted by micro-Raman mapping. The resulting fibers exhibit smooth surface, they keep uniform, bright and broadband emission, and they are capable of guiding generated light along their length, thus constituting a novel hybrid material that can be promptly utilized to build optical circuits based on miniaturized waveguides and integrated sensing regions.

## 2. Experimental Methods

MoS$_2$ NPs were produced by exfoliation of the bulk powder (Aldrich, particle size < 2 μm) by ultrasonication for 8 hours in *N*-methyl-2-pyrrolidone (NMP). Following the sonication step, centrifugation at 1000 rpm was carried out for 45 minutes to remove bulk residues and collect the supernatant. The resulting particles were inspected optically by confocal and UV-Vis spectroscopy, morphologically by scanning electron microscopy (SEM) and by transmission electron microscopy (TEM), and chemically by energy-dispersive X-ray spectrometry (EDS).

The confocal analysis was performed on vacuum-dried drops (50 μL) cast from the achieved dispersion, placed onto a glass substrate and excited with a cw laser at wavelength, $\lambda$ = 488 nm. Micro-Raman experiments were carried out with a Renishaw InVia spectrometer equipped with a confocal optical microscope and a 532 nm excitation laser. PL lifetime measurements were also performed in confocal mode by exciting light-emitting fibers by a 470 nm pulsed diode laser at 40 MHz, and collecting signals by a time-correlated single photon counting setup. Full details on Experimental Methods, including realization of electrospun hybrid fibers, are reported in the ESI file.

## 3. Results and discussion

We produce $MoS_2$ NPs by exfoliation of the bulk powder, through ultrasonication in NMP and subsequent centrifugation. The high surface tension of NMP (40 mJ m$^{-2}$), which roughly matches with the estimated surface energy of few-layered $MoS_2$ (46.5 mJ m$^{-2}$), allows a stable dispersion to be obtained.[27-30] In addition, the exfoliation process is simple and environmentally-friendly, being driven by sonochemical cavitation phenomena through formation and implosive collapse of gas bubbles, with consequent increase of local temperature and pressure changes favoring exfoliation processes and eventually leading to the formation of nano-sized particles.

Exemplary SEM images of the obtained $MoS_2$ samples, covered by a few nm metal layer to favor inspection, are shown in Fig. 1a, displaying the presence of layered flakes with transversal size up to a few micrometers, decorated with 20-80 nm particles (inset of Fig. 1a). The elemental composition of both flakes and particles, evaluated by EDS (Fig. S1 in the ESI file) is highlighted by a sharp and intense peak at 2.3 keV due to the presence of the transition-metal dichalcogenide.[31,32]

.

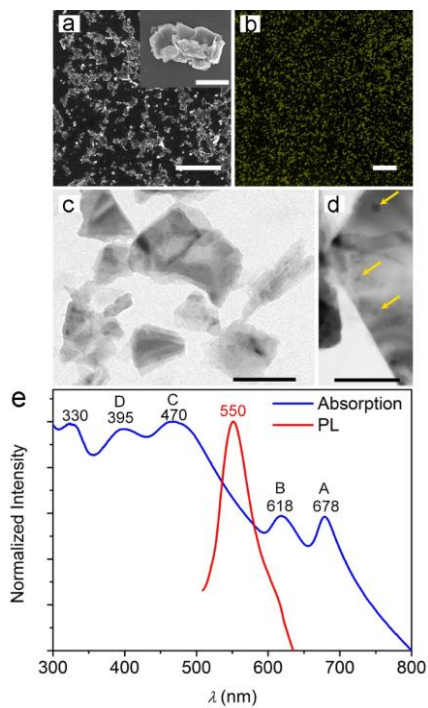

**Fig. 1.** (a) SEM micrograph of MoS$_2$ particles following exfoliation. Scale bar = 5 µm. The inset shows an exemplary single aggregate of particles at higher magnification. Scale bar = 1 µm. (b) Confocal fluorescence micrograph of a drop-cast MoS$_2$ dispersion. Scale bar = 10 µm. (c, d) TEM micrographs of MoS$_2$ following exfoliation. MoS$_2$ NPs with diameters down to 10 nm are highlighted by yellow arrows. Scale bar = 100 nm (c) and 50 nm (d), respectively. (e) Absorbance (blue line) and PL emission spectra (red line) of the MoS$_2$ dispersion.

The PL emission from cast samples appears in the form of bright spots (Fig. 1b), with maximum intensity at 550 nm and 56 nm full width at half maximum (FWHM), as typical of the fluorescence spectrum of $MoS_2$ particles with nm-scale lateral dimensions.[7,11,33] Indeed, the formation of small fragments during sonication is generally associated with a blue-shift of the emission around 450-600 nm, depending on the lateral size of the formed fragments and on the excitation wavelength. Upon excitation at 488 nm, the PL peak at 550 nm can be attributed to $MoS_2$ NPs with lateral dimension up to a few tens of nm.[7,10,11,33] This is also supported by the TEM characterization shown in Fig. 1c and Fig. 1d, which evidence the formation of few-layered $MoS_2$ flakes and highlight almost round-shaped NPs with diameter down to 10 nm.

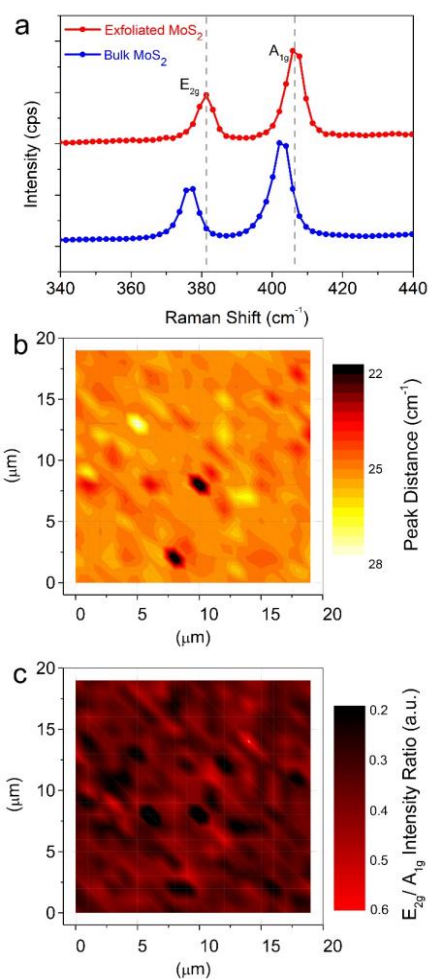

**Fig. 2.** (a) Raman spectra acquired using a 532 nm laser, for bulk $MoS_2$ (blue spectrum) and for a drop-cast $MoS_2$ dispersion after the exfoliation process (red spectrum). Spectra are vertically shifted for clarity. (b,c) Maps of the Raman mode spacing (b) and of the $E_{2g}$-$A_{1g}$ intensity ratio (c) acquired for the drop-cast dispersion (map size = 19×19 $\mu m^2$).

In order to investigate the properties of NPs more in depth we measure the UV-Vis absorption spectra of the liquid dispersion of $MoS_2$ after centrifugation. The absorbance spectrum, reported in Fig. 1e, shows characteristic features of $MoS_2$ NPs and nanosheets, with two peaks between 600 and 700 nm, and a broad band with two peaks at 395 nm and at 470 nm.[10,34] The two low-energy absorption peaks (A and B in Fig. 1e) are known to correspond to excitonic transitions, occurring between the splitted valence bands and the conductance band at the K-point of the Brillouin zone of $MoS_2$ nanosheets.[7] For instance, the measured extinction coefficient at 678 nm is 412 mL mg$^{-1}$ m$^{-1}$ (Fig. S2). Through quasiparticle self-consistent *GW* calculations, an energy difference between the splitted valence bands of 146 meV has been predicted for $MoS_2$ monolayers, and of 174 meV for bilayers, respectively, due to spin-orbit coupling and interlayer interaction.[35] Here, the energy gap of about 175 meV between the A and B transitions is therefore consistent with the formation of few-layered flakes. Additional absorption peaks at 470 nm (C in Fig. 1e), and at 395 nm (D) are regarded as being associated with transitions involving other regions with high density of state,[36-38] as from the deep valence band to the conduction band.[10,39] In their whole, peaks A-D are typical features of exfoliated $MoS_2$. Moreover, an absorption peak in the UV (at about 330 nm) can be clearly appreciated in Fig. 1e, and attributed to the formation of NPs with small size in the dispersion and consequent quantum confinement effects.[7,10,11,39] These results are collectively in agreement with previous findings suggesting how exfoliation procedures might lead to complex dispersions of heterodimensional $MoS_2$ nanostructures.

The effects of the ultrasonication process on the $MoS_2$ bulk powder can be better analyzed by Raman spectroscopy. Fig. 2a shows the comparison of the Raman spectra of the $MoS_2$ pristine powder (blue line), and of NPs after the exfoliation process (red line). Both the spectra exhibit two peaks, i.e. the $E_{2g}$ corresponding to the in-plane vibrational mode and the $A_{1g}$ related to the out-of-plane vibrational mode.

The $MoS_2$ pristine powder shows the characteristic $E_{2g}$ and $A_{1g}$ Raman modes at 376.8 cm$^{-1}$ and at 403.2 cm$^{-1}$, respectively. The corresponding vibrational modes in the exfoliated $MoS_2$ nanocrystals are shifted up to 381.6 cm$^{-1}$ and 406.8 cm$^{-1}$, respectively. The stiffening of the Raman modes is likely to be attributed to the nanostructuring of $MoS_2$ in all the three spatial dimensions following the exfoliation process. Indeed, this particular effect has been previously reported for 18 nm $MoS_2$ nanocrystals.[7] In addition, the spectral spacing between the $E_{2g}$ and $A_{1g}$ modes (25.2 cm$^{-1}$) confirms that our $MoS_2$ particles are few-layered, consisting indicatively of 6 to 10 layers.[40] The pristine powder exhibits a corresponding Raman mode spacing equal to 26.4 cm$^{-1}$ in agreement with the value of bulk $MoS_2$.[40] Interestingly, the intensity ratio between the $E_{2g}$ and $A_{1g}$, often employed as benchmark of stoichiometry and defectiveness of $MoS_2$,[41,42] is similar in both the spectra ($\cong 0.5$). This aspect is especially relevant for using the achieved dispersions to build light-emitting components, since it highlights that the exfoliation process does not cause any worsening of the material in terms of sulfur understoichiometry[42] and concentration of extended defects.[43]

In order to evaluate the obtained homogeneity of the achieved dispersion of NPs, we also carry out Raman mapping on an area of about 360 μm$^2$, following drop casting on a glass substrate. Figures 2b and 2c show the maps of the Raman mode spacing and of the $E_{2g}$-$A_{1g}$ intensity ratio, respectively. In Fig. S3 we report control maps for the pristine $MoS_2$ powder, that clearly highlights how the initial system is composed by bulk material with good homogeneity in terms of Sulphur stoichiometry. The Raman mode spacing map in Fig. 2b evidences that the system obtained after ultrasonication is instead homogeneously composed by $MoS_2$ few-layer nanocrystals, with minor bulk residues. As displayed in Fig. 2c, defective areas where the intensity ratio between the $E_{2g}$ and $A_{1g}$ modes drops significantly below 0.5, are very small within the intensity ratio map (~7% of the analysed region).

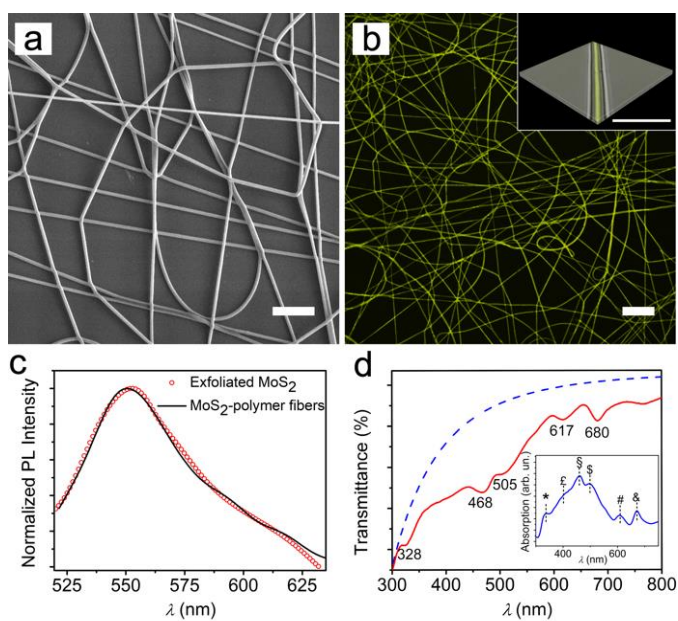

**Fig. 3.** (a) SEM micrograph of $MoS_2$-doped fibers. Scale bar = 20 μm. (b) Fluorescence confocal micrograph of fibers. Scale bar = 20 μm. The inset shows a 3D reconstruction of merged fluorescence and transmission confocal micrographs of an individual fiber by *Z*- axis scanning. Scale bar = 10 μm. (c) PL emission spectra of exfoliated $MoS_2$, drop-cast on glass (red circles) and of $MoS_2$-polymer fibers (black line). The PL peak is at about 550 nm. (d) Transmittance spectrum of aligned fibers. The blue dashed line is the calculated scattering background (~$\lambda^n$ with *n*=-3.66). Inset: scattering background-corrected absorption spectrum. Peaks are at 336 (*), 401 (£), 460 (§), 497 ($), 611 (#) and 673 nm (&).

Fibers embedding such $MoS_2$ NPs as dopants are prepared by electrospinning after dissolving the polymer matrix in the complex dispersion (see ESI for details). The resulting hybrid fibers (Fig. 3a) are uniform with average diameter around 1.6 μm and smooth surfaces (Fig. S4). Confocal fluorescence imaging also highlights a uniform PL intensity along the longitudinal axis of each fiber (Fig. 3b), thus suggesting a homogenous incorporation of $MoS_2$ in the filaments. This is further supported by the 3D reconstruction of fibers as obtained by Z-axis scanning (i.e., perpendicular to the substrate on which fibers are deposited, inset of Fig. 3b).

In Fig. 3c we compare the PL of the achieved $MoS_2$ particles with that of the fibers. No significant difference can be appreciated in the emission, which is well preserved upon electrospinning. Similarly, the absorbance spectrum (Fig. 3d) indicates that the characteristic features of $MoS_2$ are generally well preserved in the fibers, including the UV component at about 330 nm which allows one to conclude that NPs are effectively embedded in the hybrid fibrous material.

In fact, reduced transmission is also indicative of remarkable light-scattering phenomena arising in the fibrous sample, which can be straightforwardly taken into account as proportional to the *n*-power of wavelength with $-4<n<-1$[34,37] (dashed line in Fig. 3d). Once such light-scattering component is considered, fiber absorption also shows features which can be ascribed to the C (at $\cong$ 460 nm) and D ($\cong$ 400 nm) transitions of $MoS_2$ once undergone to more significant layering in the polymer matrix, together with an additional peak at 496 nm which is also likely related to aggregation of larger $MoS_2$ flakes once in the filaments (inset of Fig. 3d). In this respect, polymer fibers represent a very interesting environment in which edge and quantum effects[44] of exfoliated two-dimensional materials can be highlighted within confined volumes.

In order to study the PL lifetime of the $MoS_2$ doped fibers, we use time-correlated single photon counting detection. The retrieved PL lifetime (Fig. 4a) displays a decay with average characteristic time of (3.1 ± 0.1) ns. This result is comparable to those reported in previous works, where observed lifetimes typically ranged from 1.5 to 5.9 ns .[10,11,45-47] The lifetime map in the inset shows a homogeneous green color along all the fibers, indicating that the average lifetime is uniform in all the filaments. Worthy of note, the acquired data are better represented by fitting to a biexponential model with associated lifetimes of (1.3 ± 0.1) ns and (4.4 ± 0.3) ns and fractional contributions of 60% and 40%, respectively. The quality of the fit is assessed by the goodness of obtained residuals (monoexponential- and biexponential-derived ones are reported in Fig. 4b as green dots and as a blue line, respectively). Such biexponential behavior is not surprising, as it was already highlighted by others for $MoS_2$ NPs:[10,11] it might reasonably depend on the presence of different species of $MoS_2$ nanostructures within the fibers.[10] In fact, heterodimensional $MoS_2$ nanostructures present different lifetimes.[45]

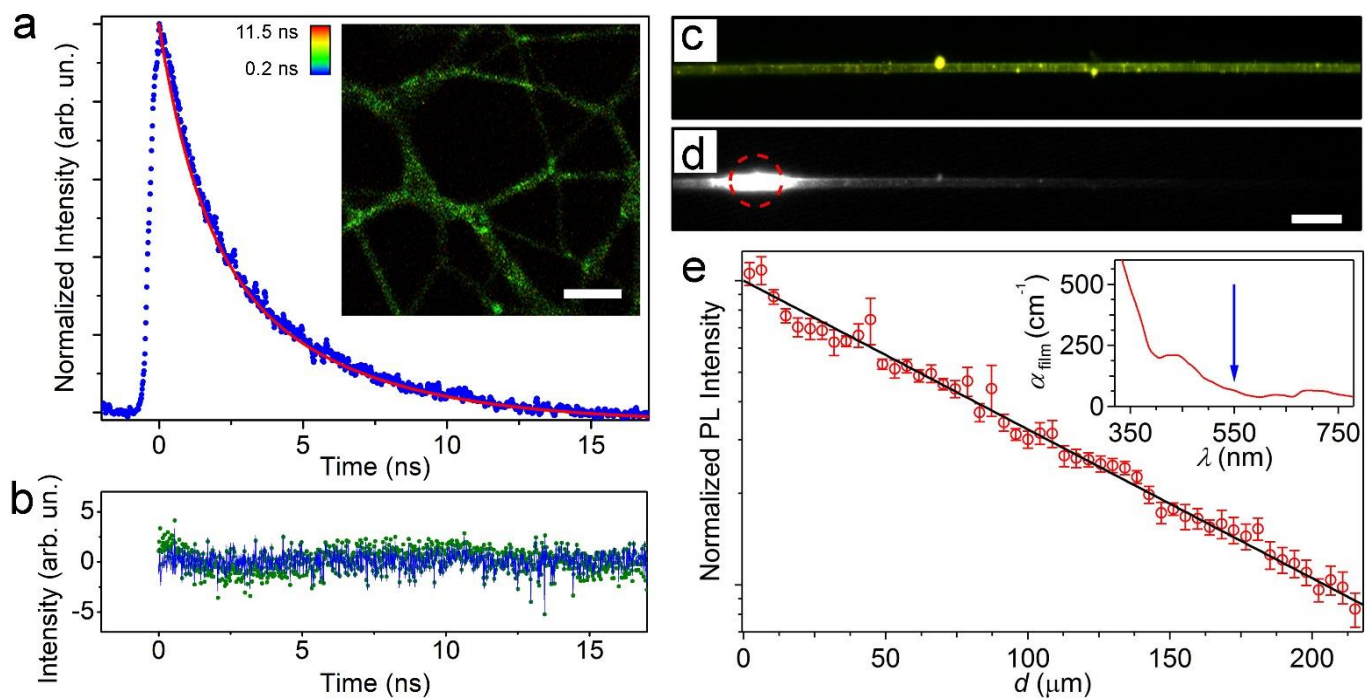

**Fig. 4.** (a) Fluorescence lifetime of $MoS_2$ doped fibers. The data are fitted using a biexponential decay model (red line). The inset shows an average lifetime map obtained by fitting the fluorescence decay curve in each pixel of the image. Scale bar = 10 μm. (b) Residuals of the monoexponential (green dots) and biexponential (blue line) fitting. (c) Fluorescence micrograph of a freestanding $MoS_2$-doped fiber. Bright spots are associated with particle clusters. (d) PL of the same fiber during excitation by a focused laser beam (red circle in the left part of the image). Scale bar = 10 μm. (e) Spatial decay of the light intensity guided inside the fiber *vs* distance, *d*, from the photoexcitation spot. The continuous line is the best fit to an exponential decay, according to the function $I = I_0^{-\alpha d}$. Inset: wavelength dependence of propagation losses in a reference film doped with $MoS_2$. The vertical arrow highlights the analyzed emission wavelength ($\cong$ 550 nm).

Finally, the waveguiding features of $MoS_2$-doped, light-emitting fibers are analyzed by microphotoluminescence (μ-PL). Suspended fibers are realized to this aim (Fig. 4c) in order to avoid any optical loss induced at the substrate/fiber interface. To assess the propagation losses for light emitted by $MoS_2$ NPs and channeled along an individual hybrid filament, the intensity of the PL escaping from the fiber surface and from the tip was imaged and measured as a function of the distance ($d$) of the tip from the excitation spot. Fig. 4d shows a typical PL image of the fiber while excited with a focused laser beam (red circle). The spatial decay of the self-waveguided emission is displayed in Fig. 4e. The continuous line is the best fit of the experimental data by an exponential function, $I_{PL} = I_0 \exp(-\alpha d)$ where $I_{PL}$ is the PL intensity, $I_0$ is a pre-exponential factor which stands for the intensity measured at very small $d$ values, and $\alpha$ is the loss coefficient. The fits lead to $\alpha = 110$ cm$^{-1}$, roughly corresponding to a light transport length of 90 µm, whose losses are significantly lower than values measured in other hybrid nanofibers such as those doped with CdSe quantum dots, and in line with the best values found in conjugated polymer nanofibers.[48,49] Theoretically, for PMMA nanofibers in air, optical losses affecting photons carried along the length of the waveguides are calculated by considering Rayleigh scattering from the nanofiber surface roughness (a few nanometers as measured by atomic force microscopy) and are expected to be in the range 10-100 cm$^{-1}$.[48,50] The optical losses for carried photons along the hybrid fiber can be also due to self-absorption by the active component (i), in addition to light-scattering from either bulk (ii) or surface (iii) defects or inhomogeneities. Instead, the contribution due to the absorption from the polymer matrix can be neglected, being of the order of $10^{-3}$-$10^{-2}$ cm$^{-1}$ in the visible range.[51]

The contributions [(i)+(ii)] due to absorption and Rayleigh scattering by bulk defects, which is mainly related to the refractive index contrast between the polymer and the inorganic fillers, can be determined by measuring losses for an optical beam propagating through a film of known thickness and with the same composition as nanofibers. The inset of Fig. 4e shows the spectrum of losses ($\alpha_{\text{film}}$) for UV-visible light passing through such reference film. These data allow one to estimate losses due to MoS$_2$ self-absorption and bulk scattering at the emission peak wavelength (about 550 nm) to be about 70 cm$^{-1}$. Additional waveguiding losses directly related to the structure and defects of MoS$_2$-polymer filaments are therefore limited to a few tens of cm$^{-1}$, indicating how uniform and free of defects these fibers are and suggesting them as highly promising building blocks for optical circuits and ports.

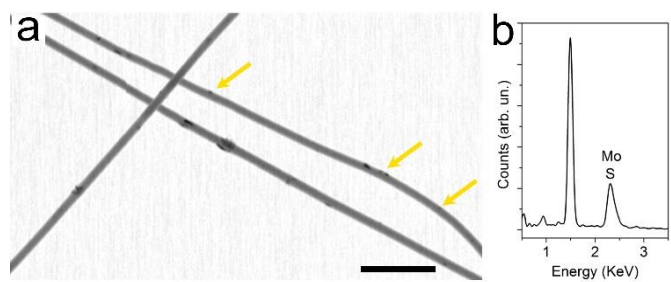

**Fig. 5.** (a) STEM micrograph of hybrid nanofibers. Interspersed particles and aggregates are clearly visible (some are highlighted by yellow arrows). Scale bar = 10 µm. The corresponding EDS profile is displayed in (b). The peak at 1.5 keV belongs to Aluminum which, in the form of tin foils, is used as collector for the deposition of the fibers.

For instance, light-scattering from inhomogeneities along the fiber axis could be promoted by clusters of particles, which can be appreciated by bright spots as those shown in Fig. 4c-d. The typical distance between consecutive light-scattering spots in µ-PL micrographs is indeed of a few µm, which well agrees with the inter-cluster distance found by STEM imaging of the hybrid fibers (Fig. 5).

**4. Conclusions**

In conclusion, we produced $MoS_2$ NPs with a simple and inexpensive ultrasonication process and realized hybrid fibers based on them. The exfoliation strategy reduces the $MoS_2$ powder size in all the three spatial dimensions, preserving low defect concentration and homogeneity in terms of Sulphur stoichiometry. $MoS_2$ is then homogeneously encapsulated in electrospun polymer filaments, which are smooth and uniform and show photonic functionality i.e. waveguiding capability of the self-emitted light along their longitudinal axis. Advantages of hybrid photonic systems based on $MoS_2$ over light-emitting nanocomposites embedding well-established semiconducting quantum dots might be numerous, particularly in terms of excitation-dependent emission features, and low cytotoxicity,[52] which make them interesting for biomedical applications including endoscopy and optogenetics. Furthermore, the large surface of these dopants can be bio-functionalized through different routes,[53,54] thus opening perspectives for use in diagnostic lab-on-chip, as well as in miniaturized chemical and optical biosensors.[55] Finally, the light transport along these $MoS_2$-polymer fibers is found to be efficient over distances of many tens of µm. Since enabling photon channeling over length-scales typical of photonic chips, $MoS_2$-polymer fibers might find application in quantum technologies,[50] and since providing electromagnetic field confinement, they can be exploited for the optical detection of field-interacting chemical species placed nearby the waveguide, or for coupling radiation to other active nanomaterials, through the evanescent field at their surface. All these aspects make these novel optical materials highly interesting for integration in complex, system-level platforms for sensing, and in flexible optoelectronics.

**Conflicts of interest**

There are no conflicts to declare.

**Acknowledgements**

The research leading to these results has received funding from the European Research Council under the European Union's Seventh Framework Programme (FP/2007-2013)/ERC Grant Agreement n. 306357 (ERC Starting Grant "NANO-JETS"). D.P. also acknowledges the support from the project PRA_2018_34 ("ANISE") from the University of Pisa.

# Low-defectiveness exfoliation of MoS$_2$ nanoparticles and their embedment in hybrid light-emitting polymer nanofibers

Alberto Portone,[a,b] Luigi Romano,[a,b] Vito Fasano,[a] Riccardo Di Corato[a,c], Andrea Camposeo,[b] Filippo Fabbri,[d] Francesco Cardarelli,[e] Dario Pisignano,[b,f] Luana Persano[b]

a. Dipartimento di Matematica e Fisica 'Ennio De Giorgi', Università del Salento, via Arnesano, I-73100 Lecce, Italy
b. NEST, Istituto Nanoscienze-CNR, Piazza San Silvestro 12, I-56127 Pisa, Italy. E-mail: luana.persano@nano.cnr.it
c. Center for Biomolecular Nanotechnologies (CBN), Istituto Italiano di Tecnologia, Via Barsanti, 73010 Arnesano (LE), Italy
d. Center for Nanotechnology Innovation @NEST, Istituto Italiano di Tecnologia, Piazza San Silvestro 12, I-56127 Pisa, Italy
e. NEST, Scuola Normale Superiore, Piazza San Silvestro 12, I-56127 Pisa, Italy
f. Dipartimento di Fisica, Università di Pisa, Largo B. Pontecorvo 3, I-56127 Pisa, Italy. E-mail: dario.pisignano@unipi.it

Electronic Supplementary Information

**Experimental Details**

*MoS$_2$ exfoliation.* Bulk MoS$_2$ powder (99%, < 2 μm in size, Sigma Aldrich) was dispersed in *N*-methyl-2-pyrrolidone (NMP) (99.5% anhydrous, Sigma Aldrich) with a concentration of 2 mg mL$^{-1}$ and sonicated for 8 hours, at room temperature, using a bath sonicator (CEIA, CP102 digit) at a constant power of 200 W. Following ultrasonication, the obtained product was centrifuged at 1000 rpm for 45 minutes and the supernatant was used for further characterization. The concentration of the dispersion was measured by completely removing the solvent. The dispersion was centrifuged at 14,000 rpm for 45 minutes until the complete sedimentation of MoS$_2$, then the transparent supernatant was collected and the residual solvent was removed by storing under vacuum for 24 h. Finally, the sediment was weighted and dispersed in NMP at known concentration.

*Electrospinning.* For realizing hybrid nanofibers, MoS$_2$ was firstly dispersed in NMP at the concentration of 3.4 mg mL$^{-1}$ and ultrasonicated for 1 hour prior electrospinning. Fibers were obtained by dissolving 350 mg of poly(methyl methacrylate) (PMMA) in 1 mL MoS$_2$/NMP dispersion, stirring and sonicating the so-achieved solution for 4 hours. At PMMA concentrations below 350 mg mL$^{-1}$, fibers were not uniform, showing beaded morphology. The electrospinning process underwent extensive optimization, varying the applied voltage in the range 10-20 kV, the needle-to-collector distance in the range 5-20 cm and the flow rate in the range 0.2-2 mL h$^{-1}$. Finally, the process was carried out by a syringe with a 21 gauge stainless steel needle, an applied voltage of 14 kV, a needle to collector distance of 15 cm and a flow rate of 0.5 mL h$^{-1}$. Both a flat plate and a rotating collector (disk with 0.8 cm width, 8 cm diameter and speed 4000 rpm) were used to obtain randomly oriented and aligned nanofibers, respectively.

*Spectroscopy.* Ultraviolet-visible (UV-Vis) absorption spectra were collected from MoS$_2$ dispersions and aligned, free-standing mats of fibers by using a Lambda 950 spectrophotometer (Perkin Elmer Inc.). Fluorescence and transmission micrographs were acquired by an inverted microscope Eclipse Ti equipped with a confocal A1R-MP system (Nikon), using an Argon ion laser (excitation wavelength, $\lambda$ = 488 nm). The sample emission was collected by a 20× (Numerical Aperture, NA = 0.50, Nikon) and a 60× (oil immersion NA = 1.40, Nikon) objective and the fluorescent signal was detected by a spectral detection unit equipped with a multi-anode photomultiplier (Nikon). Confocal microscopy was performed on both dried drop-casted MoS$_2$ dispersions and fibers, deposited on glass substrates. A 3D reconstruction of fibers was obtained in Z-stack mode with a step size of 0.3 μm.

Micro-Raman experiments were carried out with a Renishaw InVia spectrometer equipped with a confocal optical microscope and a 532 nm excitation laser. The spectral resolution of the system is 1 cm$^{-1}$. The micro-Raman spectra was acquired with a 50× objective (NA 0.8), a laser power of 1 mW, and an acquisition time of 2 s. The peak fitting was carried out with a Levenberg-Marquardt algorithm, employing a Voigt function for the evaluation of the peak parameters.

*Morphological characterization.* Morphological analysis of exfoliated MoS$_2$ and of electrospun fibers was performed by scanning electron microscopy (SEM) and scanning transmission electron microscopy (STEM, FEI Nova NanoSEM 450). The microscopy was equipped with an energy dispersive X-ray (EDS) detector and operating at acceleration voltages of 6-20 kV. Samples of exfoliated MoS$_2$ were prepared by drop-casting on a Si surface. All the samples were metallized by thermal evaporation (chromium or aluminum, PVD75, Kurt J. Lesker Co.) prior to SEM analysis. Transmission electron microscopy (TEM) images were acquired with a JEOL Jem1011 microscope operating at an accelerating voltage of 100 kV. For this analysis, few drops of MoS$_2$ dispersion were drop-casted onto carbon-coated copper grid (300 mesh). TEM images were collected and processed by using the Gatan Microscopy Suite (GSM) software.

*Confocal fluorescence lifetime imaging microscopy (FLIM).* PL lifetime measurements were performed by an inverted microscope with confocal head (TCS SP5, Leica Microsystem) and a 40× objective (NA=1.24). A 470-nm pulsed diode laser operating at 40 MHz was used as excitation source. The fluorescence intensity from light-emitting fibers was collected in the 480-600 nm range by a photomultiplier tube interfaced with a time-correlated single photon counting setup (PicoHarp 300, PicoQuant, Berlin). FLIM acquisitions lasted until an average of $10^2$-$10^3$ photons were collected in each pixel.

*Waveguiding.* Waveguiding was studied by using a micro-photoluminescence system, based on an inverted microscope (IX71, Olympus) equipped with a 20× objective (NA = 0.50, Olympus) and a charge coupled device (CCD) detector. The sample photoluminescence was excited by a continuous wave diode laser ($\lambda$ = 405 nm), coupled to the microscope by a dichroic mirror and focused on the sample through the microscope objective (spot diameter about 10 μm). Individual freestanding fibers were suspended on a holder with a slot of 0.5 cm. The laser beam was focused on a single emissive fiber and fluorescence images were acquired using the CCD camera. After the excitation, part of the light emitted by MoS$_2$ was coupled into the fiber and then guided. Therefore, using the acquired images, the optical losses were evaluated by measuring the intensity of the light scattered from fiber surface as a function of the distance, *d*, from the exciting laser spot.

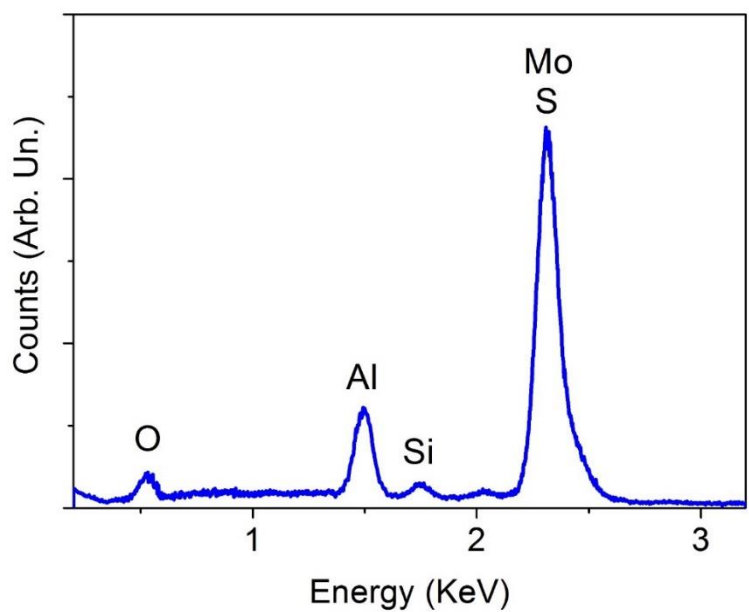

**Fig. S1.** EDS profile of drop-casted MoS$_2$ dispersion. Silicon belongs to the substrate on which fibers are deposited, while aluminum was thermally evaporated onto the fibers for SEM imaging.

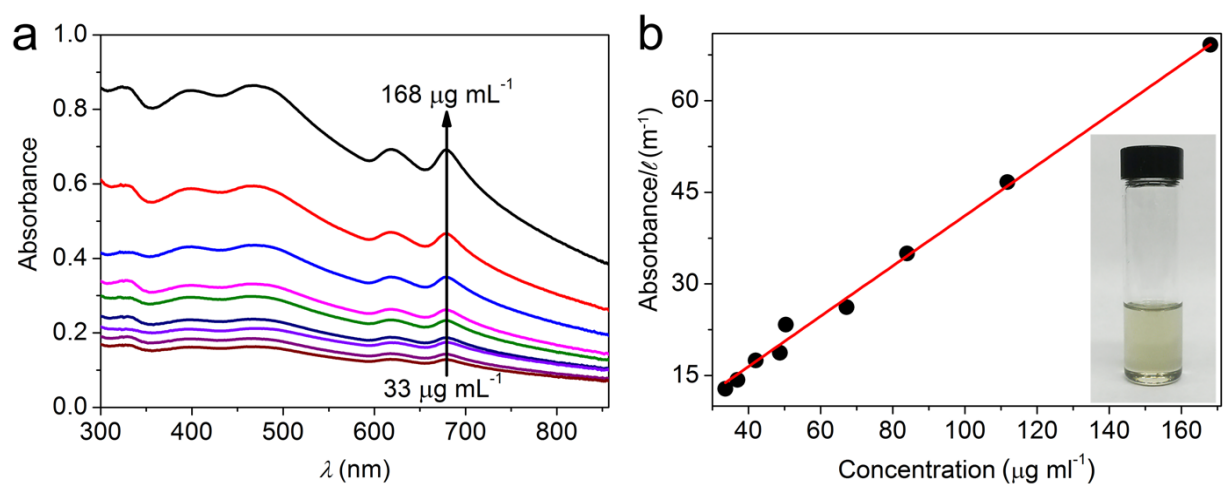

**Fig. S2**. (a) Absorbance spectra of MoS$_2$ dispersions at different concentration. From bottom to top, concentration = 33 µg mL$^{-1}$, 37 µg mL$^{-1}$, 42 µg mL$^{-1}$, 48 µg mL$^{-1}$, 50 µg mL$^{-1}$, 67 µg mL$^{-1}$, 84 µg mL$^{-1}$, 111 µg mL$^{-1}$, 168 µg mL$^{-1}$. (b) Lambert-Beer linear plot for MoS$_2$ dispersions. The slope of the line provides the extinction coefficient at $\lambda_{678\,nm}$: 412 mL mg$^{-1}$ m$^{-1}$. l is the optical path length. The inset shows a MoS$_2$ dispersion in NMP (1:10) with a concentration of 84 µg mL$^{-1}$.

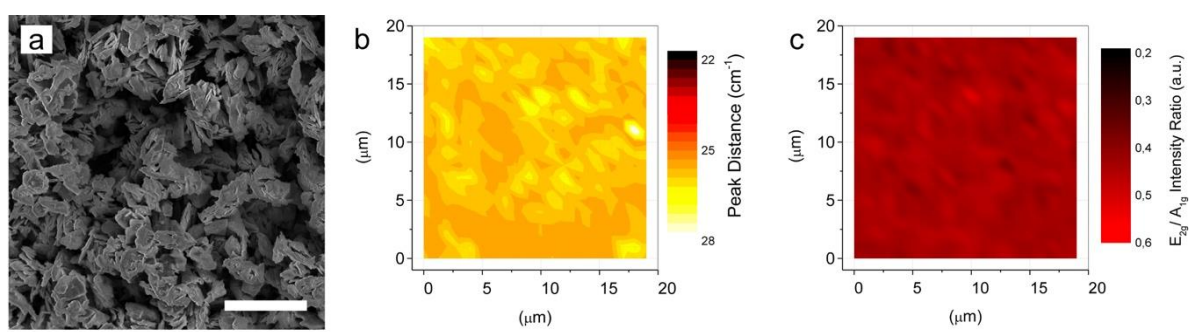

**Fig. S3.** (a) SEM micrograph of the bulk MoS$_2$ powder. Scale bar = 5 μm. Maps of the Raman mode spacing (b) and of the E$_{2g}$-A$_{1g}$ intensity ratio (c) acquired on bulk MoS$_2$.

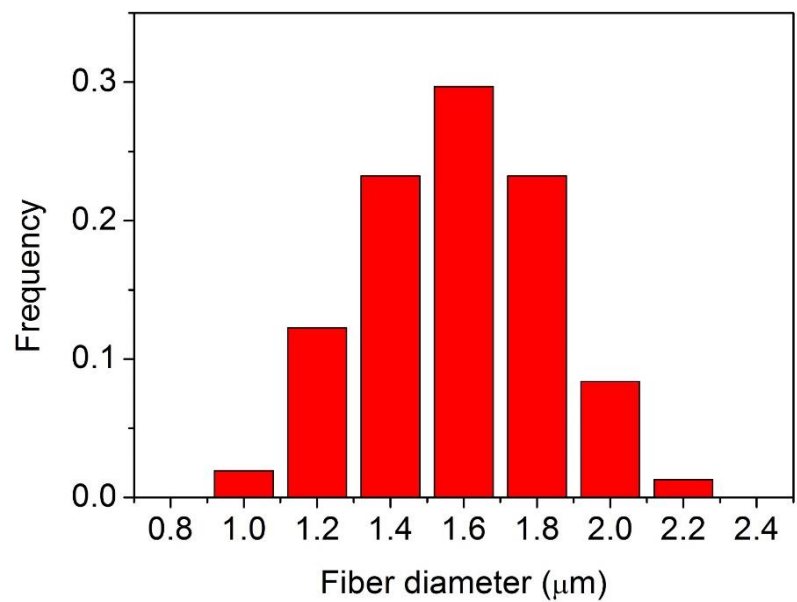

**Fig. S4.** Diameter distribution of the hybrid $MoS_2$-polymer fibers.